\documentclass[aps,prc,twocolumn,showpacs]{revtex4}
\usepackage{graphicx} 
\usepackage{dcolumn} 
\usepackage{bm}         
\begin{document}

\title{A Hydrodynamic and Transport Model for Ultrarelativistic Heavy
Ion Collisions at RHIC and LHC Energies}

\author{
Yu-Liang Yan$^1$, Yun Cheng$^2$, Dai-Mei Zhou$^2$, Bao-Guo Dong$^{1}$, Xu Cai$^2$, Ben-Hao Sa$^1$, and Laszlo P. Csernai$^{3,4}$}

\affiliation{
$^1$
China Institute of Atomic Energy,
P. O. Box 275 (10), 102413 Beijing, China\\
$^2$
Institute of Particle Physics,
Huazhong Normal University,
430079 Wuhan, China\\
$^3$
Department of Physics and Technology,
University of Bergen, 5007, Bergen, Norway\\
$^4$
MTA-KFKI Research Inst for Particle and Nuclear Physics,
1525 Budapest, Pf. 49, Hungary
}
\date{\today}

\begin{abstract}

Combining the hydrodynamic model (Hydro code) and the transport model (PACIAE model), we present the Hydro-PACIAE hybrid model. We use the Hydro-PACIAE hybrid model to calculate Au+Au collisions at $\sqrt{s_{\rm{NN}}}$=130 GeV and Pb+Pb collisions at $\sqrt{s_{\rm{NN}}}$=2.76 TeV. The obtained pseudo-rapidity and transverse momentum distributions well reproduce the experimental data. This shows that the Hydro-PACIAE hybrid model is useful to describe the ultrarelativistic Heavy Ion Collisions at RHIC and LHC Energies. We used the hybrid model to calculate the elliptical flow for the Pb+Pb collisions at $\sqrt{s_{\rm{NN}}}$=2.76 TeV. The $v_2$ values are bigger than the experimental data at high $p_T$, and will become bigger when the Hydro evolution time is longer. The results indicate that the selection of the transition hypersurface has significant effect on the observable quantities.

\end{abstract}


\maketitle

\section{ Introduction }
The nucleus-nucleus collisions at the Relativistic Heavy Ion Collider (RHIC) produced an initial hot and dense matter and this has been interpreted as a strongly coupled quark-gluon plasma (sQGP) \cite{brah,phob,star2,phen}. The Large Hadron Collider (LHC) opens up opportunities to explore the quark gluon plasma (QGP) in a wider energy region. The hydrodynamic models are useful tools to describe the hot and dense matter generated in the early stage of ultrarelativistic heavy ion collisions. Here, interactions are strong and frequent, so that the microscopic models that assume binary collisions with free propagation of constituents between collisions have limited validity. The advantage of hydrodynamic models is that one can vary flexibly the dissipative properties and the equation of state (EoS) of the matter and test its consequences on the reaction dynamics and the outcome. On the other hand, the initial and final freeze-out (FO) stages of the reaction are outside
the domain of applicability of the hydrodynamic model.

Microscopic models, e.g. the transport model PACIAE, are quite successful to describe the partonic and hadronic stages of the collision. Unfortunately,
the transport models are usually restricted to binary collisions. Thus their validity is limited in the hot and dense matter especially when the QGP is generated. Furthermore, explaining the QGP phase transition between the hadronic and the partonic phase on a microscopic level is also an open issue that still has to be resolved.

Recently the so-called hybrid models, combining microscopic and macroscopic modules, have been developed. There are several hybrid models \cite{Petersen,Bass,Hirano,Bauchle,Abelev}. In this paper, we present a hybrid model, Hydro-PACIAE, which combines the hydrodynamic model (Hydro code) and the transport model (PACIAE model). The hybrid model connects the different stages of the reaction and each of these are matched \cite{match} appropriately. The initial state for hydrodynamic model is constructed based on the so-called Effective String Rope model \cite{Magas, Magas2}, and considering that the system is close to local equilibrium at high energies. The subsequent fluid dynamical model assumes a QGP fluid in local equilibrium. Then we consider a transition from the QGP fluid phase to the partonic/hadronic phase, which is described by a combination of a fluid dynamical (FD) and a particle transport model, PACIAE. Here we try to gain more insight into the principles, which may govern these processes.

We use the hybrid model, Hydro-PACIAE, first to simulate the final hadron distribution in pseudo-rapidity and transverse momentum, as well as comparing the model results with the experimental data for Au+Au collisions at $\sqrt{s_{\rm{NN}}}$=130 GeV and Pb+Pb collisions at $\sqrt{s_{\rm{NN}}}$=2.76 TeV. Then we use the hybrid model to calculate the elliptic flow in Pb+Pb collisions at $\sqrt{s_{\rm{NN}}}$=2.76 TeV.

\vskip -1mm

\section{Hybrid Model Description}
The most essential observables in ultrarelativistic heavy ion reactions
\cite{ALICE-Flow1} from the laboratory detector can help us to understand the properties of the dense and hot matter. In general, we expect to extract more information from experiments, both on the Equation of State (EoS) and the transport properties of matter \cite{Son,CKM}. We have to apply a realistic and suitable description with fully 3+1 dimensional dynamical evolution at all stages of the reaction.

\subsection{hydrodynamic Model for initialization in Hybrid model}

The hydrodynamic stage is calculated by using the Particle In Cells (PIC) method \cite{PIC1957}, which uses a mesh of cells with smaller, Lagrangian fluid elements in them. The model assumes the MIT Bag Model equation of state and relativistic perfect fluid. The total number of so called ``Marker particles" (which are the Lagrangian cells) is about 5 million, and their number remains a constant during a collision. Marker particles are defined to carry the same amount of conserved baryon charge, thus this charge is also exactly conserved by the method. Each cell of the calculational mesh contains a larger number of marker particles, placed in the mesh randomly. The marker particles move among the cells of the mesh during the fluid dynamical development. As we know, hydrodynamic models assuming phase and chemical equilibrium during all stages of the reaction cannot account for all observations in general. However, they provide an efficient way to describe the nearly perfect fluid phase of QGP. Then, we transform the fluid in the cells into a set of partons or hadrons, which are then described with the PACIAE model.

The quark-gluon fluid is assumed to have a small viscosity \cite{Son}, especially in the vicinity of the critical point of the phase transition to/from hadronic matter \cite{CKM}. This phase transition \cite{CK92} has significant dynamical effect on the hadronization, that in local, fluid dynamical phase equilibrium takes a long time, which contradicts to observations. To avoid this time delay recently rapid transition scenarios are assumed, which can be handled well in the hybrid approach.

The 3+1 dimensional, relativistic FD model, we use to describe energetic
heavy ion reactions, is well established and describes the measured
collective flow phenomena well \cite{hydro1,hydro2}.
In our Computational Fluid Dynamics (CFD) calculation
the initial state model is based on expanding flux tubes or streaks
\cite{Magas, Magas2}. It is a system where the matter is stopped within each
streak, while streaks expand independently of each other.
Thus, this model is applicable streak by streak and the
momentum of the streaks varies, especially for the peripheral streaks
where the asymmetry between the projectile and target contribution
to the participant matter is the biggest. So, the streaks at the
projectile and target sides move in the beam ($z-$) direction with
substantial velocity difference. In the initial state model \cite{Magas, Magas2} the initial transverse velocity is zero for all fluid elements.

In the CFD simulations with the PIC solution method \cite{hydro1,hydro2}, the equations of relativistic FD were solved for a perfect quark-gluon fluid.
At the same time the numerical method, due to the finite grid
resolution, led to dissipation and thus to entropy production,
which was analyzed \cite{Horvat}. From the entropy production
we could determine the corresponding "numerical viscosity",
and this was approximately the same as the estimated, low viscosity
of the quark gluon plasma. To avoid double counting, i.e. over
counting, of viscous dissipation, we did not add additional viscous
terms to our CFD model simulations.

In case of heavy ion reactions the flow is not stationary
and the shear flow geometry is only present in the initial
state.  Later, due to the large pressure of QGP the plasma
explodes and expands almost radially in a way that the final flow pattern
at freeze out is close to spherical, somewhat elongated longitudinally
and in the reaction plane ($\pm x$-direction) due to the dominant
elliptic flow.

\subsection{Basic Description in PACIAE model}

The PACIAE model \cite{Yan, Cheng, Sa} is a parton and hadron cascade model, which is based on PYTHIA \cite{PYTHIA}. The model consists of four stages: The parton initialization, which decomposes the fluid into partons; The post transition partonic state consists of quarks, anti-quarks, and gluons; Then, in the parton evolution stage, we are evaluating the total and the differential cross sections arising from the partonic evolution, which can be simulated by the Monte Carlo method until the parton-parton collisions are ceased; Finally, the partons reach the hadronization at the moment of freeze-out (FO). This is described by the phenomenological fragmentation model (String fragmentation) and the coalescence model \cite{Coalescence1, Coalescence2}. After hadronization, the rescattering produces the final hadrons, and we obtain a configuration of hadrons in the spatial and momentum coordinates after the hadronization. The detailed description for PACIAE is the following:

\begin{enumerate}
\item PARTON INITIALIZATION: In the case if we start directly from two colliding nuclei, an A+A (A+B) collision is decomposed into binary nucleon-nucleon (NN) collisions based on the collision geometry, so that every collision time is calculated by
assuming that the nucleons propagate along straight line trajectories
and interact with the NN inelastic (total) cross sections. Then the
initial NN collision-list is constructed from these NN collisions.
A NN collision with the earliest collision time is selected
from the collision list, and the final state of the collision is obtained
by the PYTHIA model with string fragmentation switched-off. Afterwards
the diquarks (anti-diquarks) are broken randomly into quark pairs
(anti-quark pairs), and one obtains a configuration of quarks,
anti-quarks, and gluons, beside a few hadronic remnants for a NN collision.

In case of the hybrid model, we generate partons from the fluid cells of the FD model at a transition hyper-surface. This is described later.

\item PARTON EVOLUTION:  The evolution is made via parton rescattering, with $2 \rightarrow 2$, binary LO-pQCD differential cross
sections \cite{BL1977}. Then the total and differential cross sections in the
parton evolution (parton rescattering) are simulated
by the Monte Carlo method.

\item HADRONIZATION: The parton evolution stage is
followed by the hadronization at the moment of partonic
freeze-out. This means no more parton collisions. In the PACIAE model, the Lund string fragmentation model and
phenomenological coalescence model are supplied for the
hadronization of partons after rescattering. In case of the hybrid model the latter one is used.

\item After hadronization the re\-scattering among
produced hadrons is dealt with the usual two-body collision model.
Only the rescatterings among $\pi, K, p, n, \rho (\omega), \Delta,
\Lambda, \Sigma, \Xi, \Omega, J/\Psi$ and their antiparticles are
considered in the calculations. An isospin averaged parametrization
formula is used for the $hh$ cross section \cite{koch,bald}. We also
provide an option for the constant total, elastic, and inelastic
cross sections ($\sigma_{\rm{tot}}^{NN}=40$~mb, $\sigma_{\rm{tot}
}^{\pi N}=25$~mb, $\sigma_{\rm{tot}}^{kN}=35$~mb,
$\sigma_{\rm{tot}}^{\pi \pi}=10$~mb) and the assumed ratio of
inelastic to total cross section of 0.85.

\end{enumerate}

\subsection{Connection of Stages in Hybrid Model}

We present the connection of the hydrodynamic model and PACIAE model, as the transfer of the fluid cell by cell to partons, satisfying the conservation laws based on the assumption of the thermal equilibrium before and after the transition. The detailed steps are as follows:

$(1)$ The hydrodynamic model, assumes a QGP fluid, with an EoS given by the MIT Bag model with massless quarks and gluons and a background field. Each Lagrangian fluid cell has a given baryon charge, which are rescaling the charge density in the cells of the mesh,  $n(x,y,z)$. We may deal with the transition hyper-surface in $3$ optional ways: The first case is the simplest assumption that transition takes place isochronously, $i.e.$ at $t=const.$; The second case is that the transition happens on a hyperboloid, $i.e.$ at $\tau=\sqrt{t^2-\vec{r}^2}=const.$, where $\tau$ is the proper transition time from a space-time point. The third case is when the transition is described as a sudden change across a space-time hyper-surface, which hyper-surface is calculated based on a condition (e.g. a transition temperature), if the transition condition is known. To be simple, in this work, we choose the first case, in which the direction of the flow and normal vectors are everywhere the same, and they are perpendicular to the transition hyper-surface.

$(2)$ We use the perfect relativistic fluid dynamical equations:
$$ N^\mu_{,~\mu} = 0 ~~~~ {\rm and}~~~~T^{\mu\nu}_{,~\mu} = 0.$$
where $u^\mu = (\gamma, \gamma \vec{v})$, and $ w = e +P $, $ T^{ik} = w
\gamma^2 v_i v_k + P \delta_{ik}$. During the transition, the above dynamical equations can be simplified as conservation laws:
\begin{equation}
  [ T^{\mu\nu} d\hat{\sigma}_\nu ] = 0,~~~  [ N^\mu d\hat{\sigma}_\mu ] = 0.\label{ConserveEqs}
\end{equation}

In our simulation, the above conservation laws have to be satisfied \cite{match} and the examination of the increasing entropy should be done during the whole simulation. For example, baryon charge, $N_i$, and the energy, $E_i$,  are calculated according to the formula for each cell on the transition hyper-surface:
\begin{eqnarray}
N_i=V^{LR}_{cell}~n_i &=&V^{LR}_{cell}~N_{i}^{\mu}u_{\mu}, \\
E_i=V^{LR}_{cell}~e_i &=& V^{LR}_{cell}~u_{\mu}T_{i}^{\mu\nu}u_{\nu}.
\end{eqnarray}
where the index $``i" $ runs over the different generated parton or hadron species.  $V^{LR}_{cell}$ is the volume of the cell in the local rest frame. $u_{\mu}$ is the post transition 4-velocity, $u^{\mu}=\gamma(1,\vec{v})$, and $u_{\mu}=\gamma(1,-\vec{v})$, where $\gamma=1/\sqrt{1-\vec{v}^2}$. $\vec{v}$ is the 3-velocity of the fluid cell in the partonic or hadronic phase. Note that this velocity after the transition is not the same as the flow velocity of the QGP. The energy momentum tensor $T_{i}^{\mu\nu}$ includes the contribution of the particles with their kinetic energies. In the post transition stage, particles in partonic matter include quarks, anti-quarks and massless gluons. The detailes of this transition are described in the Appendix \ref{appendix}.

$(3)$ We utilize the Stefan Boltzmann distribution to generate the partons, or J\"{u}ttner distribution to generate the hadrons, considering the strangeness suppression and mass difference in the phase transition. We give the ratio of generated parton types in equilibrium state with considering the quark masses via the strangeness suppression factor $e.g. \gamma_s=0.3$, according to the mass difference of the light quarks and strange quarks at an average transition temperature. Of course, this ratio also affects the momentum distribution for different parton types.

$(4)$ The 4-momenta of the particles are generated according to the Cooper-Frye formula \cite{cooper}. The baryon chemical potentials for the given baryon number densities are taken into account. As we mentioned above, we are choosing a time step ($t=Constant$), and dispersing the partons into the fluid cells with proper distributions after the hyper-surface.

These steps are done for all fluid cells of the hydro code, which are carrying baryon number density, pressure and momentum. The created partons serve as parton initialization for the PACIAE model.

\section{Assessment in Ultrarelativistic Heavy Ion Collisions}\label{Asses}
\begin{figure}
\vskip -1mm
 \centering
  \includegraphics[width=3.0in]{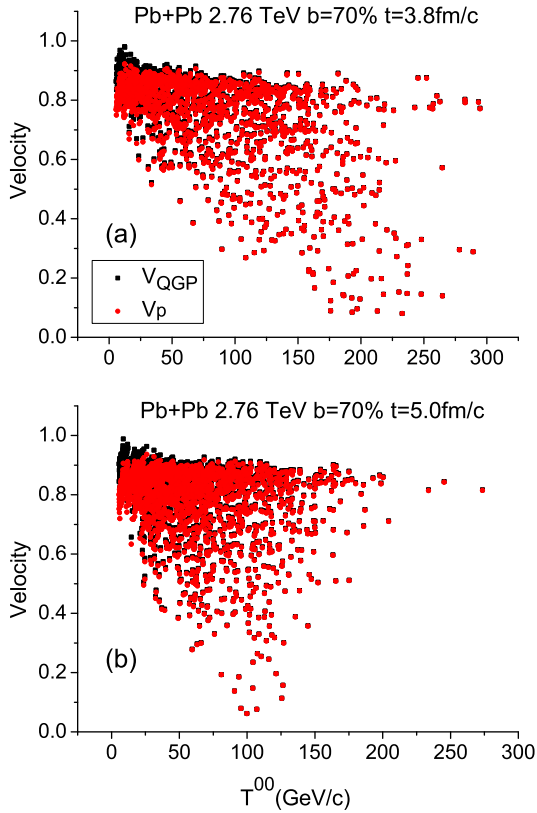}
\vskip -1mm
 \caption{
Velocity change in the
transition as function of the energy density in different transition
time steps, (a) 3.8 fm/c and (b) 5.0 fm/c, in the Hydro-PACIAE hybrid model calculation. The cell by cell changes from the QGP (black squares) to the partonic matter
(red dots) are modest except at low energy densities where the velocity
visibly decreases.}
 \label{v_ts138}
\end{figure}

\begin{figure}
\vskip -1mm
 \centering
  \includegraphics[width=3.0in]{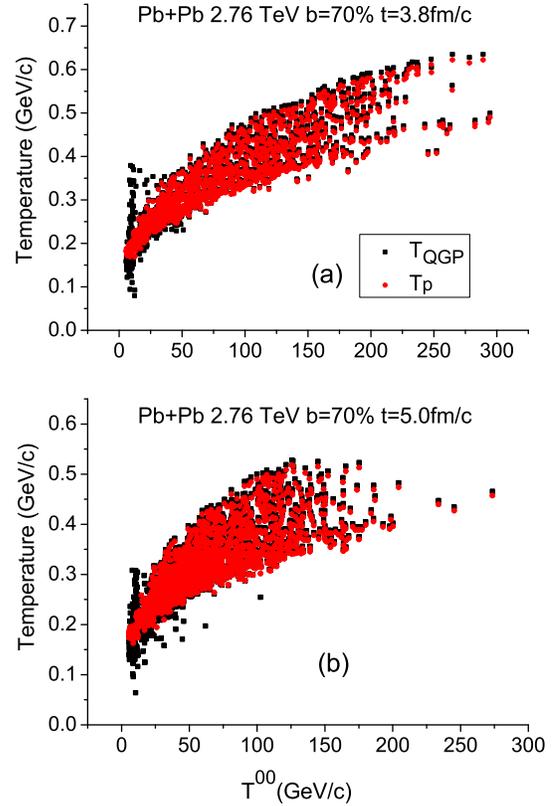}
\vskip -1mm
 \caption{
Temperature change  in the
transition as function of the energy density in different transition
time steps, (a) 3.8 fm/c and (b) 5.0 fm/c, in the  hybrid model calculation. The cell by cell changes from the QGP (black squares) to the partonic matter
(red dots) are modest except at low energy densities where the temperature
slightly increases and becomes more concentrated.}
 \label{T_ts138}
\end{figure}

Figures \ref{v_ts138} and \ref{T_ts138} show the velocity and temperature change in the transition from the QGP fluid to partonic matter as function of the energy density for different transition time steps, 3.8 fm/c and 5.0 fm/c, in the Hydro-PACIAE hybrid model calculation. We can see that the cell by cell changes from the QGP fluid to partonic matter are modest except at low energy densities where the velocity and temperature change visibly.

\begin{figure}[h]
\vskip -1mm
 \centering
  \includegraphics[width=3.4in]{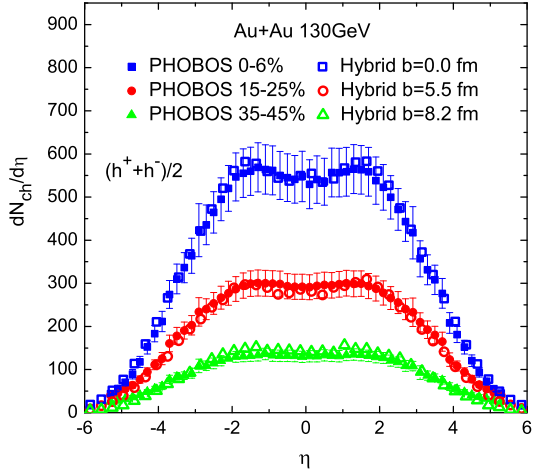}
\vskip -1mm
 \caption{
Charged particle pseudo-rapidity distribution from the Hydro-PACIAE hybrid
model calculations, and PHOBOS data \cite{PHOBOS} in Au+Au collisions at $\sqrt{s_{\rm{NN}}}$=130 GeV.
The experimental data are given in the indicated centrality
bins (full symbols) and for these the corresponding mean impact parameters
(open symbols) are indicated for the hybrid model calculations.}
 \label{F_130_eta}
\end{figure}

For comparing the results of Hydro-PACIAE hybrid model with the results of Hydro code conveniently, we calculated the Au+Au collisions at $\sqrt{s_{\rm{NN}}}$=130 GeV. The final charged particle pseudo-rapidity distribution is shown in Fig. \ref{F_130_eta}, where the calculated results of Hydro-PACIAE model is compared with the PHOBOS data \cite{PHOBOS}. The observed centrality bins correspond to a set of impact parameters, which yield to different multiplicities depending on the impact parameter and the FO time. This is illustrated in Ref. \cite{hydro2} for the case of Pb+Pb reactions at LHC energy, and the dependence is similar at lower energies also. In the FD model the calculated charged particle multiplicity, $N_{ch}$, is a function of FO time (assuming a $t_{FO} = const.$ FO hyper-surface), and of the impact parameter, $b$. These multiplicities can matched to the measured mean charged multiplicity in a given experimental multiplicity bin. For example, the impact parameter $b=5.5$ fm for Au+Au collisions corresponds approximately to the experimental centrality of 20\% and $b=8.2$ fm for centrality of 40\%. This is similar to the impact parameter method in transport model \cite{sa2002}.

\begin{figure}[h]
\vskip -1mm
 \centering
  \includegraphics[width=3.4in]{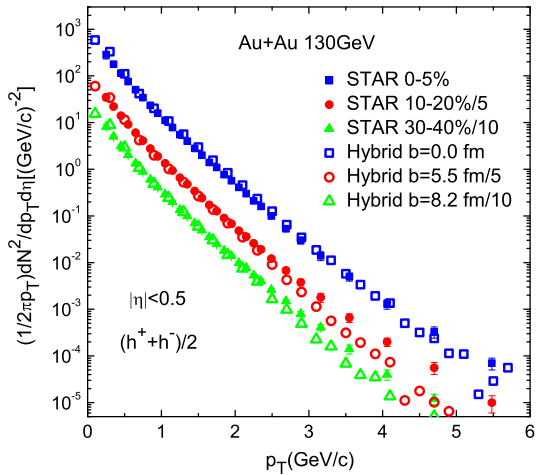}
\vskip -1mm
 \caption{
$p_T$ distribution for the Hydro-PACIAE hybrid model calculation and STAR data \cite{star} in Au+Au collisions at $\sqrt{s_{\rm{NN}}}$=130 GeV, for the same centrality bins and mean impact parameters
as in Figure \ref{F_130_eta}. }
 \label{F_130_pt}
\end{figure}

We also calculated the transverse momentum distribution for the Au+Au collisions at $\sqrt{s_{\rm{NN}}}$=130 GeV and compared the results with the STAR data\cite{star} in Fig. \ref{F_130_pt}. The good agreement in Figs. \ref{F_130_eta} and \ref{F_130_pt} indicate that our Hydro-PACIAE hybrid model can successfully reproduce the experiment data at RHIC energy.

\begin{figure}[h]
\vskip -1mm
 \centering
  \includegraphics[width=3.4in]{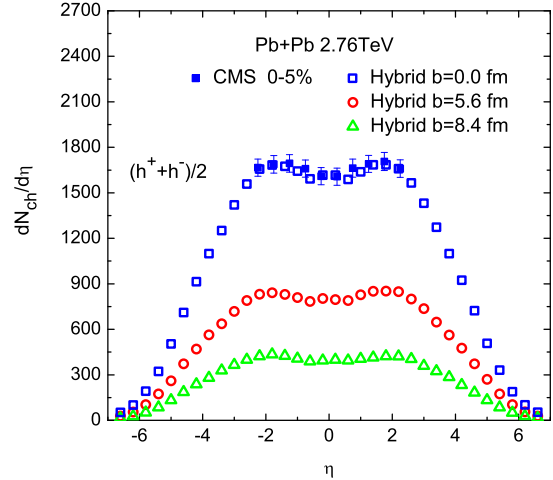}
\vskip -1mm
 \caption{
Charged particle pseudo-rapidity distribution from the Hydro-PACIAE hybrid
model calculations, and CMS data \cite{cms} in central collisions.
The experimental data are given in the indicated centrality
bin (full symbols) and the mean impact parameters
(open symbols) are indicated for the hybrid model calculations.}
 \label{F_2760_eta}
\end{figure}

\begin{figure}[h]
\vskip -1mm
 \centering
  \includegraphics[width=3.4in]{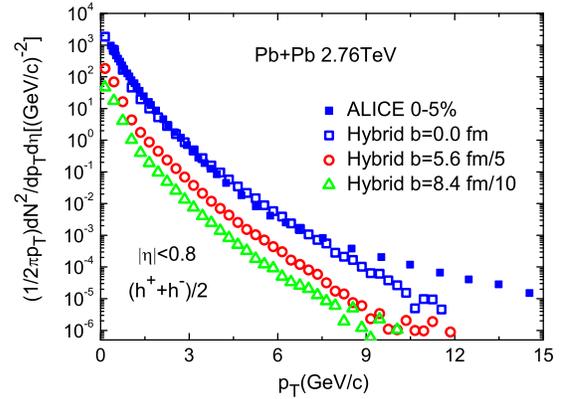}
\vskip -1mm
 \caption{
$p_T$ distribution for the Hydro-PACIAE hybrid model calculation,
and ALICE data \cite{alice} in central collisions, for the same centrality
bin and mean impact parameters as in Figure \ref{F_2760_eta}. }
 \label{F_2760_pt}
\end{figure}

Figure \ref{F_2760_eta} shows the pseudo-rapidity distribution of Hydro-PACIAE hybrid model for the Pb+Pb collisions at $\sqrt{s_{\rm{NN}}}$=2.76 TeV together with the CMS data, which are taken form \cite{cms} and the pseudo-rapidity range is from -2.5 to 2.5. We can see that in the $|\eta|<2.5$ area, the hybrid model results are well compared with the CMS data. The transverse momentum distribution for the Pb+Pb collisions at $\sqrt{s_{\rm{NN}}}$=2.76 TeV are also calculated and the results are shown in Fig. \ref{F_2760_pt} compared with the ALICE data \cite{alice}. We can see that in the low and middle $p_T$ regions the hybrid model results well reproduce the ALICE data, but the model results are smaller in the high $p_T$ region.

\begin{figure}[h]
\vskip -1mm
 \centering
  \includegraphics[width=3.4in]{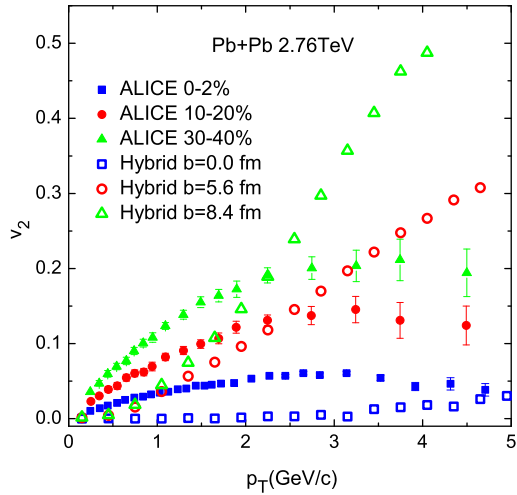}
\vskip -1mm
 \caption{
$v_2$ distribution for the Hydro-PACIAE hybrid model calculation, and ALICE
data in central collisions and semi central collisions, for the same
centrality bins and mean impact parameters as in Figure \ref{F_2760_eta}.
The 0-2\% ALICE data are taken from \cite{alice2}, and 10-20\%, 30-40\%
are taken from \cite{alice3}.}
 \label{F_v2}
\end{figure}

\begin{figure}[h]
\vskip -1mm
 \centering
  \includegraphics[width=3.4in]{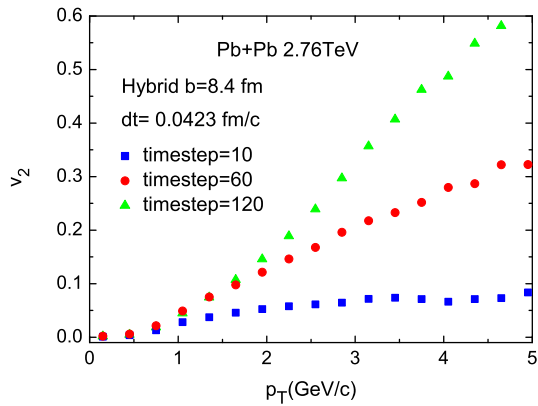}
\vskip -1mm
 \caption{
$v_2$ distribution for the Hydro-PACIAE hybrid model calculation with different time step evolution. }
 \label{F_v2_step}
\end{figure}

We used the Hydro-PACIAE hybrid model to calculate the elliptical flow for the Pb+Pb collisions at $\sqrt{s_{\rm{NN}}}$=2.76 TeV. The results are compared with the ALICE data\cite{alice2, alice3} in Fig. \ref{F_v2}.We see that in the low $p_T$ region the Hydro-PACIAE hybrid model results are smaller than the ALICE data, but the model results are bigger than the ALICE data for the high $p_T$ region. This feature originates from the hydro model \cite{hydro2}, mainly because the Hydro code in the Hydro-PACIAE hybrid model is ideal fluid dynamics and has little viscosity. The FO time is also an important quantity for the elliptical flow. We compared the $v_2$ with different FO time evolution for 40\% centrality Pb+Pb collisions at $\sqrt{s_{\rm{NN}}}$=2.76 TeV in Fig. \ref{F_v2_step}. The $v_2$ value becomes bigger when the Hydro evolution time increases from 0.423 to 2.54 and 5.08 fm/c. This is because when the Hydro evolution time is longer more elliptic flow develops with time in the Hydro-PACIAE hybrid model.

The directed flow, $v_1(\eta)$ for Pb+Pb collisions at $\sqrt{s_{\rm{NN}}}$=2.76 TeV was also evaluated in the hybrid model, for centrality percentage bins
40\% and 60\%. The obtained $v_1(\eta)$ showed a random statistical fluctuation around zero with a $v_1(\eta)$-amplitude of 0.1\% for $ | \eta | < 4$, which
increased to 0.2\% for $ | \eta | > 4$, due to the smaller charged particle
multiplicity at higher pseudorapidities. The obtained $v_1(p_T)$ was zero at
low $p_T$ and reached to 1\% (3\%) at $p_T = 3 (4)$GeV/c respectively due to
statistical fluctuations with decreasing charged particle multiplicity.
Thus, the fluctuations arising from the transport model PACIAE, dominated the
smaller azimuthal asymmetry, $v_1(\eta)$, from the initial FD stage of the
hybrid model. The same situation was observed for $v_1(p_T)$, where the
FD expectation for global collective flow vanishes due to symmetry reasons.

The elliptic flow, $v_2(\eta)$ was also dominated by strong statistical
fluctuations.  Both for the directed and elliptic flow the reaction plane
of the evaluation was the theoretical reaction plane given by the
impact parameter vector. The initial states, arising from the FD stage
of the model did not include any random fluctuation. These results are
consistent with the experimental finding that the central and semi-central
collisions are dominated by random fluctuations, especially by random
initial state fluctuations, which are not correlated with the reaction
plane.

\section{Conclusions}
\vskip -1mm
Combining the hydrodynamic model (Hydro code) and the transport model (PACIAE model), we have presented the Hydro-PACIAE hybrid model. We used the Hydro-PACIAE hybrid model to calculate Au+Au collisions at $\sqrt{s_{\rm{NN}}}$=130 GeV and Pb+Pb collisions at $\sqrt{s_{\rm{NN}}}$=2.76 TeV. The pseudo-rapidity and transverse momentum distributions well reproduce the experimental data. This shows that the Hydro-PACIAE hybrid model is effective to describe the ultrarelativistic Heavy Ion Collisions at RHIC and LHC Energies. We used the hybrid model to calculate the elliptical flow for the Pb+Pb collisions at $\sqrt{s_{\rm{NN}}}$=2.76 TeV. The $v_2$ values are bigger than the experimental data in the high $p_T$ region, and it shows that the $v_2$ value will become bigger when the Hydro evolution time is longer. This result indicates significant dependence on the choice of the transition surface between the two components of the hybrid model. Furthermore, the space-time shape of the FO surface may also influence the final anisotropy produced by the model.

The fluid dynamical model needs an initial stage of relativistic heavy ion collisions to start, and the so-called Effective String Rope model was used to generate the initial stage for Hydro code \cite{Magas, Magas2}. We also plan to use the transport model, PACIAE, to generate the initial stage for Hydro code instead of the Effective String Rope model. Then we will present a full transport model with an intermediate hydrodynamic stage to study the relativistic heavy ion collisions. The variability of the model components, will enable us a bigger flexibility to study the role of different model assumptions, initial state and transition conditions on a wide set of observables.

\section{Appendix}\label{appendix}
\subsection{The transition procedure}

Here we evaluate and present the details of the transition from the FD stage
to the parton cascade stage of the model in case of a $t=Constant$ transition
hypersurface.
All components of the energy momentum tensor of the QGP are described in
the PIC hydrodynamic model. The hydro code uses the Bag Model EoS with
$N_F=2$. However, some of the fluid cells with low internal energy would have
negative pressures due to the Bag constant, which are set to zero in the model.
Thus, we need to calculate the conserved quantities in the transition from QGP
to partonic matter both for the cases of positive and zero pressure. First,
in the QGP phase, the energy density component and momentum current component,
as well as the baryon charge current are as follows:
\begin{eqnarray}
T^{00}_{QGP}&=&(e_{QGP}+P_{QGP})\gamma^2_{QGP}-P_{QGP}=e_{cf}\label{t00q}\\
T^{10}_{QGP}&=&(e_{QGP}+P_{QGP})\gamma^2_{QGP}v_{QGP}\label{t10q}\\
N_{QGP}^0&=&n_{QGP}\gamma_{QGP}
\end{eqnarray}
where $e_{cf}$ is the energy density in the QGP phase in the calculational
frame (CF). In the general case, when the 3-velocity of the fluid is directed
arbitrarily, $\vec{v}_{QGP}=(v_{QGP-x},v_{QGP-y},v_{QGP-z})$, which is known
from the hydro code, then the parton side flow will keep the velocity
{\it 3-direction} in each cell because we assumed that the transition happens
at a $t=Constant$ hyper-surface. Then after transition the velocity
relationship between the pre- and post- transition velocities can be written as:
$$v_x^P=\frac{v_x^{QGP}}{|\vec{v}_{QGP}|}v_P,
~~~v_y^P=\frac{v_y^{QGP}}{|\vec{v}_{QGP}|}v_P,
~~~v_z^P=\frac{v_z^{QGP}}{|\vec{v}_{QGP}|}v_P$$
Here $|\vec{v}_{QGP}|=\sqrt{v_x^2+v_y^2+v_z^2}$ in the QGP phase and
the r.h.s quantities are known from hydro code. On the parton side, we have
$$
\gamma_P^2=\frac{1}{1-\vec{v}_P^2},~~{\rm or}~~
v_P=\sqrt{\frac{\gamma_P^2-1}{\gamma_P^2}} \ ,
$$
and we use the EoS in the QGP phase:
$$
e_{QGP}+P_{QGP}=4(P_{QGP}+B_{bag})
$$

\paragraph{\bf Positive pressure in fluid cells:}
When the fluid cells have positive pressure, based on Eqs.
(\ref{t00q}), (\ref{t10q}) and using the EoS, we can obtain $P_{QGP}$
and $e_{QGP}$, which are expressed by the known $e_{cf}$, and
$v_{QGP}$ ($\gamma_{QGP}$) as:
\begin{eqnarray}
P_{QGP}&=&
\frac{e_{cf}-4B_{bag}\gamma^2_{QGP}}{4\gamma^2_{QGP}-1}\label{pqform}\\
e_{QGP}&=&
\frac{3e_{cf}+4B_{bag}\gamma^2_{QGP}v_{QGP}^2}{4\gamma^2_{QGP}-1} \ .
\end{eqnarray}
Inserting these into the Eqs. (\ref{t00q}) and (\ref{t10q}), we get the
conserved quantities $T^{00}$ and $T^{10}$
in terms of $e_{cf}$ and $v_{QGP}$ as:
\begin{eqnarray*}
T^{00}_{QGP}&=& e_{cf}\\
T^{10}_{QGP}&=&
\frac{4\gamma^2_{QGP}v_{QGP}}{4\gamma^2_{QGP}-1} (e_{cf}-B_{bag})
\end{eqnarray*}

\paragraph{\bf Zero pressure in fluid cells:}
In this case, we use the general definition, Eqs. (\ref{t00q}) and (\ref{t10q}) with the condition $P_{QGP}=0$ for these fluid cells, and we obtain the simple
expressions:
\begin{eqnarray*}
T^{00}_{QGP}&=& e_{cf}\\
T^{10}_{QGP}&=& e_{cf}v_{QGP} \ .
\end{eqnarray*}
This is necessary because in these cells with zero pressure, we do not use
the MIT Bag model EoS, and the relation between $T^{00}$ and $T^{10}$
should be consistent with the modification.

\subsection{Solving the conservation laws}

The transition from QGP to partonic phase have to satisfy the conservation
laws, Eq. (\ref{ConserveEqs}), and the solution of this boundary condition
is done in \cite{match}.
Here we defined the conserved energy momentum current,
${\bf A}^{\mu}$,
$$
{\bf A}^\mu \equiv  T^{\mu\nu} d\sigma_\nu~ =~w~
u^\mu u^\nu d\sigma_\nu - P g^{\mu\nu}d\sigma_\nu~,
$$
and the baryon current $j$,
\begin{equation}
j\equiv N^\mu d\sigma_\mu=n~ u^\mu d\sigma_\mu~.\label{jcurrent}
\end{equation}
With the notation, $w= e + P$, the above two quantities are a 4-vectors and
an invariant scalar, which can be considered as crossing the hyper-surface
element, moreover, we construct two invariant scalar equations of the
conserved energy momentum current by
(i) taking its norm,  ${\bf A}^\mu {\bf A}_\mu$, and
(ii) taking its projection to the normal direction,
${\bf A}^\mu d\sigma_\mu$, they have the expressions as:
\begin{eqnarray}
{\bf A}^\mu {\bf A}_\mu
& = &
w (e-P) (u^\mu d\sigma_\mu )^2 + P^2 (d\sigma^\mu d\sigma_\mu ) \ ,
\label{AA} \\
{\bf A}^\mu d\sigma_\mu & = &
w (u^\mu d\sigma_\mu )^2 - P (d\sigma^\mu d\sigma_\mu ) \ .
\label{Ads}
\end{eqnarray}
Eliminating the term $ w (u^\mu d\sigma_\mu )^2 $ from Eqs. (\ref{AA})
and (\ref{Ads}), we obtain that
\begin{equation}
{\bf A}^\mu {\bf A}_\mu  =
(e-P) {\bf A}^\mu d\sigma_\mu + e\;P\; (d\sigma^\mu d\sigma_\mu ) ~~~~.
\label{AA2}
\end{equation}
In the same way, put Eq. (\ref{jcurrent}) into Eq. (\ref{Ads}), then we get
\begin{equation}
j^2 \frac{w}{n^2} = {\bf A}^\mu d\sigma_\mu + P (d\sigma^\mu d\sigma_\mu )\label{AJ}
\end{equation}

Thus if we know the pre transition $T^{\mu\nu}_{QGP}$ and $d\sigma_\mu$,
we can get the solution from Eqs. (\ref{AA2}) and (\ref{AJ}) to obtain the
post transition configuration, with the equation of state. For example,
using the simplest case in our simulation, where the post transition EoS is
simply $P_P=E_P/3$, then from Eq. (\ref{AA2}) we get,
\begin{equation}
d\sigma^\mu d\sigma_\mu\, E_P^2 +
2\, {\bf A}^\mu d\sigma_\mu~ E_P -
3\, {\bf A}^\mu {\bf A}_\mu = 0 \, ,\label{AAEP}
\end{equation}
and if we assume
$d\sigma^{\mu}=(1,0,0,0)$ and $u^\mu= \gamma (1, v_{QGP}, 0, 0)$,
then
$$
{\bf A}^\mu {\bf A}_\mu =
(T^{00}_{QGP})^2-(T^{10}_{QGP})^2 ~~~\rm and
~~~{\bf A}^\mu d\sigma_\mu=T^{00}_{QGP} ,
$$
and then $E_P$ can be solved from the quadratic Eq. (\ref{AAEP}) with the help of Eqs. (\ref{t00q}) and (\ref{t10q}). In the same way, we can solve $n_P$ from Eq. (\ref{jcurrent}), together with Eq. (\ref{AJ}).

\section*{Acknowledgements}\label{Ack}

This work was supported by the National Natural Science Foundation of China
under Grants No. 11105227, 11175070, 11075217, 11047142, 10975062 and the 111
project of the foreign expert bureau of China.
Dr. Yuliang Yan and Yun Cheng thank for the kind hospitality from Prof. L.P Csernai of the Institute for physics and technology, University of Bergen, where part of this work was done.

\end{document}